\documentclass[twocolumn,showpacs,preprintnumbers,amsmath,amssymb]{revtex4}

\usepackage{graphicx}
\usepackage{dcolumn}
\usepackage{bm}

\begin{document}


\title{A relativistic partially electromagnetic planar plasma shock}

\author{M. E. Dieckmann\textsuperscript{1,2}, P. K. Shukla\textsuperscript{1} 
and L. O. C. Drury\textsuperscript{2}}
\affiliation{\textsuperscript{1}Institute of Theoretical Physics IV 
and Centre for Plasma Science and Astrophysics, Ruhr-University Bochum, 
D-44780 Bochum, Germany \\
\textsuperscript{2}Dublin Institute for Advanced Studies, 5 Merrion Square,
Dublin 2, Ireland} 

\date{\today}

\begin{abstract}
We model relativistically colliding plasma by PIC simulations in one and 
two spatial dimensions, taking an ion-to-electron mass ratio of 400. Energy 
dissipation by a wave precursor of mixed polarity and different densities 
of the colliding plasma slabs results in a relativistic forward shock 
forming on millisecond timescales. The forward shock accelerates electrons 
to ultrarelativistic energies and reflects upstream ions, which drag the 
electrons along to preserve the plasma quasi-neutrality. No reverse shock 
forms. The shock may be representative for internal gamma ray burst shocks. 
\end{abstract}


\maketitle
Accreting compact objects like neutron stars or black holes are amongst 
the most energetic environments in the universe. Such systems can range 
from microquasars through active galactic nuclei to gamma ray bursts 
(GRBs) \cite{Piran,Piran2,Waxman,Zensus}. Common to these objects 
is the ejection of a relativistic jet that is probably accelerated by 
magnetic fields \cite{Bland,Nishikawa3}. GRB jets are ejected by a 
forming compact object \cite{Nature}. Their high Lorentz factor $100 < 
\Gamma < 10^3$ introduces a radiation beaming, making GRBs detectable at 
cosmological distances. The GRB emissions are explained either by the 
fireball model, in which the plasma kinetic energy dominates 
\cite{Piran,Piran2}, or by the electromagnetic model, in which the 
Poynting flux is dominant \cite{EM,Lyutikov}. We consider the 
fireball model.

The high energies involved in the thermalisation of relativistic jets imply 
that plasma structures form, such as phase space holes \cite{Hole}. A shock 
description needs a kinetic approach \cite{Fahr}. Particle-in-cell (PIC) 
simulations have modelled the external GRB shock between the jet and the 
ambient plasma \cite{Nishikawa1,Silva1,Silva2,Hededal,Frederiksen,Nishi3}, 
and have examined the magnetic field amplification and particle acceleration. 
The weakly magnetized external GRB shocks develop out of a broad wave 
spectrum \cite{Medvedev,Bret1,Bret2,Tautz}, and are thus filamentary. The 
internal shocks, powering the prompt emissions, form closer to the collapsing 
star with its strong magnetic fields. The magnetic field pressure may 
initially exceed the thermal pressure of the jet plasma, in contrast to 
the case considered by Ref. \cite{Medvedev}. The magnetic field and the 
high plasma temperature \cite{Ryde} suppress the filamentation modes 
\cite{Bret3,Nishikawa2,Dieck2} and other wave modes can develop. 

The finite grid instability requires a high spatial resolution on electron
timescales of multi-dimensional PIC simulations \cite{Dieck1}, while a 
relativistic shock will extend over ion scales. A large ion-to-electron 
mass ratio is required to model shocks \cite{Scholer}. To account for 
these constraints, we select the initial conditions such as to physically 
confine the plasma dynamics to one spatial dimension (1D) and three 
momentum components, which allows us to restrain also the simulation 
geometry to one spatial dimension. We verify the validity of this 
assumption with the help of a two-dimensional (2D) simulation. We 
demonstrate that the magnetic field results in the formation of a 
relativistic planar shock. The energy dissipation is provided 
by a wave with mixed electrostatic and electromagnetic polarity, probably
a nonlinear oblique whistler \cite{Oblique}. This contrasts the energy
dissipation by the purely electromagnetic filamentation modes considered
in Refs. \cite{Silva1,Silva2,Hededal,Frederiksen,Nishikawa2,Nishi3}, and 
the nonrelativistic electrostatic shocks modelled in Refs. 
\cite{Scholer,Sorasio}. Different slab densities \cite{Sorasio} facilitate 
the development of a shock with a Mach number of 40 and an Alfv\'enic Mach 
number of 10. The forward shock reflects upstream ions.

PIC simulation methods \cite{Eastwood} solve the Maxwell equations 
and the relativistic equation of motion
\begin{eqnarray}
\nabla \times \mathbf{E} = -\frac{\partial \mathbf{B}}{\partial t} \, , \\
\nabla \times \mathbf{B} = \mu_0 \mathbf{j} + \mu_0 \epsilon_0
\frac{\partial \mathbf{E}}{\partial t} \, ,\\
\frac{d \mathbf{p}_i}{dt} = \frac{q_j}{m_j} \left ( \mathbf{E}+\mathbf{v}_i
\times \mathbf{B} \right ) \, ,
\end{eqnarray}
and fulfill $\nabla \cdot \mathbf{B}=0$ and $\nabla \cdot \mathbf{E}=\rho
/ \epsilon_0$ as constraints. The electric $\mathbf{E}$ and magnetic 
$\mathbf{B}$ fields are defined on a discrete grid. Each computational 
particle (CP) with index $i$ of the species $j$ has the charge $q_j$ and 
mass $m_j$ with $q_j / m_j = q / m$, where $q$ and $m$ are the physical 
charge and mass of the particles, represented by the species $j$. The 
particle momentum is $\mathbf{p}_i =m_i\mathbf{v}_i \Gamma (\mathbf{v}_i )$.  

Two plasma slabs collide in the simulations at the position $x=0$, each 
consisting of the electrons and ions. The ion-to-electron mass ratio is 
$R=m_i /m_e = 400$. The plasma frequency of the species $j$ with the number 
density $n_j$ is $\omega_{p,j} = {( n_j q^2_j / m_j \epsilon_0 )}^{1/2}$. 
The electrons of slab 1 have $\omega_{p,1} = 10^5 \times 2\pi / s$ and the 
ions $\omega_{p,2}= \omega_{p,1}/ \sqrt{R}$. The dense slab 1 is initially 
moving with the positive speed $v_b$ along the $x-$direction. The electrons 
and ions of slab 2 have $\omega_{p,3} = \omega_{p,1} / \sqrt{10}$ and 
$\omega_{p,4} = \omega_{p,3} / \sqrt{R}$, respectively. The tenuous slab 2 
moves with the speed $-v_b$ along the $x-$direction. Initially, all mean
speeds along the $y,z$ directions vanish. All plasma species have the 
observed jet temperature $T = 100$ keV \cite{Ryde}. The position and time 
have the units of the electron skin depth $\lambda_s =c /{( \omega_{p,1}^2 
+ \omega_{p,3}^2 ) }^{1/2}$ and $\lambda_s / c$, where $c$ is the speed of 
light in vacuum. Each slab initially occupies half of the simulation 
box. The slabs collide along the $x-$direction at the speed $v_c = 2 v_b 
/ (1+v_b^2) = 0.9 \, c$, which is representative of internal GRB shocks 
\cite{Piran2}. We introduce a spatially homogeneous oblique magnetic field. 
The magnetic field component with $e B_x / m_e = \sqrt{2} \omega_{p,1}$ 
suppresses the filamentation instability \cite{Bret3} and a perpendicular 
$B_z = B_x / 10$ enhances the energy dissipation. Initially, we set 
$B_y = 0$. The initial magnetic field strength is thus $5 \, \mu T$. The 
initial box-averaged plasma kinetic energy density exceeds the magnetic 
field energy density by a factor of $10^4$. The 1D simulation box length is 
$L = 1.58 \times 10^4 \lambda_s$, which is resolved by $1.8 \times 10^5$ 
cells. The 2D box has the extent $L_x=714 \, \lambda_s$ ($1.6 \times 10^4$ 
cells) and $L_y = 11 \, \lambda_s$ ($200$ cells) along the $x$ and $y$ 
directions, respectively. The boundary conditions are periodic. The 1D (2D) 
simulation time is $t_1 = 4370$ ($t_2 = 194$). The 2D simulation represents 
each plasma species by 64 particles per cell (PPC). The 1D simulation 
resolves each species of the dense (tenuous) slab by 576 (256) PPC.

\begin{figure}
\includegraphics[width=8.2cm]{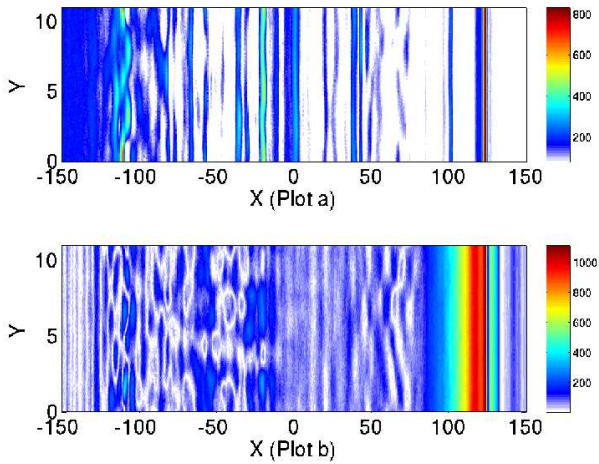}
\caption{\label{fig1} (Color online) The electric fields in V/m in the 2D 
simulation at $t=t_2$: (a) shows the flow-aligned electric field component 
$E_x$ and (b) shows $E_y$, the second electric field component in the 
simulation plane.}
\end{figure}

The electric field in the 2D simulation plane at $t=t_2$ in Fig. \ref{fig1} 
reveals a planar structure at $x \approx 120$, the leading edge of the 
dense plasma slab. A similar $E_z$ distribution is not shown. The 
filamentation instability would lead to the structure formation on a scale 
of $\lambda_s$ along $y$, and it is here suppressed by $B_x$ \cite{Bret3}. 
No strong field is present at $x \approx -120$.
We exploit the planarity of the slab front in Fig. \ref{fig1} and resort 
to the 1D simulation with its better statistical plasma representation. 

\begin{figure}
\includegraphics[width=8.2cm]{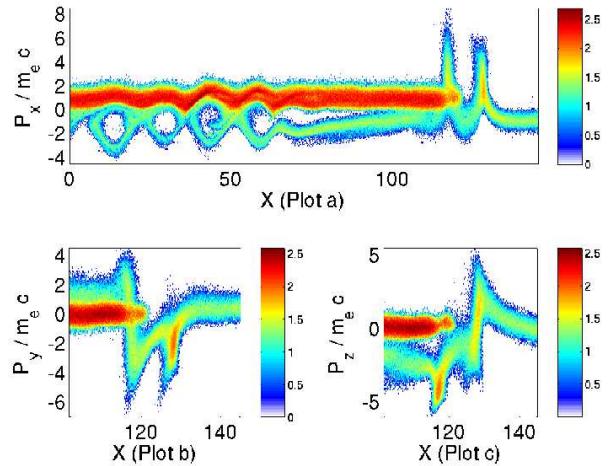}
\caption{\label{fig2} (Color online) The 10-logarithms of the electron
distributions in the 1D simulation at $t=t_2$: The distribution in the 
$x,p_x$ plane is shown in panel (a). Panels (b) and (c) display the 
distributions in the $x,p_y$ and $x,p_z$ planes.}
\end{figure}

The current system responsible for the electric fields in Fig. \ref{fig1} 
is displayed in Fig. \ref{fig2}. 
The electrons of the tenuous slab are deflected by the oblique magnetic 
field that is frozen into the dense downstream plasma. Since the massive
upstream ions are not significantly deflected, a current and thus the
electric field is building up at the leading edge of the plasma slab in 
Fig. \ref{fig1}. Further downstream, Fig. \ref{fig2}(a) shows the electron 
phase space holes, which thermalize the electrons, as in Ref. \cite{Dieck2}.
These are also observed at the Earth bow shock \cite{Bale}. 

\begin{figure}
\includegraphics[width=8.2cm]{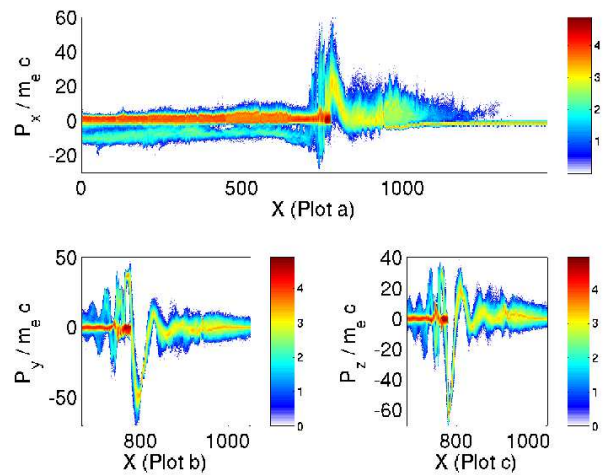}
\caption{\label{fig3} (Color online) The 10-logarithms of the electron
distributions in the 1D simulation at $t=1550$: Panel (a) shows the 
distribution $x,p_x$. The distributions in $x,p_y$ and $x,p_z$ are 
exhibited in panels (b) and (c), respectively.}
\end{figure}

\begin{figure}
\includegraphics[width=8.2cm]{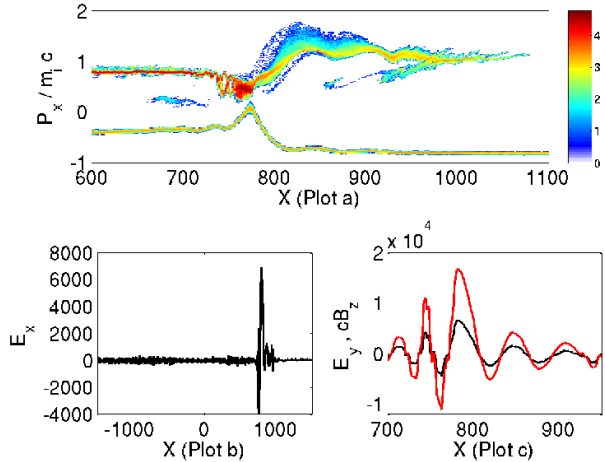}
\caption{\label{fig4} (Color online) Panel (a) shows the ion 
distribution in $x,p_x$ space. The color is the 10-logarithm of the 
number of CPs. Panel (b) shows the electrostatic $E_x$ field and panel 
(c) shows $E_y$ and $c B_z$ (red). The time is $t=1550$.}
\end{figure}

The energy dissipation by the electric fields eventually modulates the 
electrons of the dense plasma slab, giving rise to larger electric 
fields. Consequently, Fig. \ref{fig3} demonstrates that the
electrons are accelerated to $\Gamma$ factors of tens to hundreds and 
follow a corkscrew orbit. The relativistic mass of the electrons is 
not small compared to $m_i = 400 \, m_e$ at $t=1550$. 
Figure \ref{fig4}(a) shows the ion response to the strong fields.
The spatial interval with the strongest particle acceleration at $x 
\approx 780$ is well behind the expected slab front $v_b \, t /\lambda_s 
\approx 980$ at $t=1550$. Some ions have been accelerated beyond the
initial velocity $v_b$ by the electric field at the slab front 
\cite{Dieck2} and they have propagated farther upstream beyond 
$x \approx 980$. 

The wave modes at $x \approx 800$ that grow only close to the leading 
edge of the dense plasma slab, as Fig. \ref{fig4}(b) exemplifies for the 
$E_x$ field component, could have originated from an oblique whistler 
instability, similar to that discussed in Ref. \cite{Oblique}. The
wave growth at $x < 0$ is suppressed or delayed. This will introduce
an asymmetric plasma dynamics in the intervals separated by $x = 0$. 

The wave amplitudes in the considered plasma scale linearly with 
$\omega_{p,1}$ if all ratios between $\omega_{p,j}$ remain the same 
as discussed, for example, in Ref. \cite{Wakefield}. The wave is 
circularly polarized, which explains the electron corkscrew orbits 
in Fig. \ref{fig3}. The peak value of $B_z$ in Fig. \ref{fig4}(c) 
exceeds the initial background magnetic field $B_0 ={(B_x^2+B_z^2)}^{1/2}$ 
with $c B_0 \approx 1500$ V/m, and the wave in Fig. \ref{fig4}(c) is 
nonlinear. The mixed polarity of the waves allows them to interact with 
the plasma through their electrostatic component. The wave amplitude is 
sustained, because the wave is driven by the dense plasma slab that 
compensates damping losses. The damping may explain why $E_y$ and $B_z$ 
are in phase in Fig. \ref{fig4}(c). 

\begin{figure}
\includegraphics[width=8.2cm]{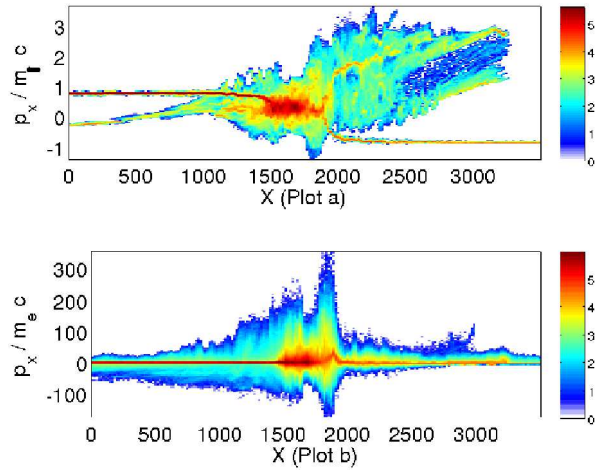}
\caption{\label{fig5} The shock: (a) shows the ion distribution in
$x,p_x$, while (b) shows the corresponding electron distribution. 
The color is the 10-logarithm of the number of CPs.}
\end{figure}

At $t=t_1$, the ion distribution in Fig. \ref{fig5}(a) has developed 
into a shock.
The foreshock in the interval $2000 < x < 3200$ is formed by incoming 
and shock-reflected ions. The electron transport with the 
shock-reflected ions across the upstream $\mathbf{B}$ in Fig. 
\ref{fig5}(b) ensures the plasma quasi-neutrality. The two ion beams at 
$x < 500$ result from the delayed shock formation. At $x \approx 1800$, 
the $p_x$ values of the electrons are comparable to those of the ions 
multiplied by $m_i / m_e$ in the simulation reference frame. A Lorentz
transformation of the electron velocity into the upstream frame, which
moves with $v = -v_b$ relative to the simulation box frame, yields
a peak electron $\Gamma \approx 700$ for our $v_c = 0.9 \, c$. In
comparison, the filamentation instability in Ref. \cite{Nishi3} yields 
only a peak electron $\Gamma \approx 10$ for a collision speed $v_f$ 
with $\Gamma (v_f) = 5$. The planar electromagnetic shock considered
here is a stronger electron accelerator. 

The shock in Fig. \ref{fig5} has formed during the time $t=t_1$, which 
is 7 ms in physical units. If we assume that this shock would constitute 
an internal shock of a GRB and that the jet frame moves at a $\Gamma 
\approx 10^3$, the time dilation would yield a duration of 
seconds for the shock development in the Earth (observer) frame of 
reference. This is comparable to the timescale of the prompt emissions.
The shock we consider is likely to form during a longer time for $m_i 
/ m_e = 1836$. The bulk Lorentz factor of GRB jets may, however, also 
be less than $10^3$ \cite{Waxman}. Different jet plasma densities, which 
may be higher than we consider in this work \cite{Meszaros}, do not 
strongly affect the shock development time due to a scaling of the 
timescales with the square root of the density \cite{Wakefield}. 

In summary, we have examined the collision of two plasma slabs with 
different densities at a relativistic speed. The bandwidth required to 
model the plasma dynamics at a high ion-to-electron mass ratio cannot 
be realized at present with PIC simulations in more than one spatial 
dimension due to the necessary large box size \cite{Waxman} and the 
high-resolution grids that are required in higher dimensions to 
suppress the electromagnetic finite grid instability \cite{Dieck1}. 
The quasi-parallel magnetic field has, however, suppressed the beam 
filamentation, and the initial planar shock development may thus be 
well-approximated by a one-dimensional PIC simulation. The energy 
dissipation has been accomplished through a nonlinear wave of mixed
polarity, which could have originated from an oblique whistler instability. 
Whistlers are frequently observed also at slower quasi-parallel plasma 
shocks and they are known to be efficient particle accelerators 
\cite{Vol}. The large energy dissipation, which ultimately forces 
the shock to form, is likely to distort the guide magnetic field and the 
slab boundary would likely be deformed by the ram pressure excerted by 
the upstream plasma. However, a significant distortion of the guide 
magnetic field away from the flow-aligned direction introduces a stronger 
perpendicular magnetic field component. The latter will probably not 
suppress or delay the shock formation and the plasma thermalization. Our 
simulation may thus give a correct estimate for the shock formation time
even in 3D.

The shock we consider here may thus be representative for a GRB internal 
shock. Such a rapid shock formation has previously been questioned, at 
least for purely electromagnetic or electrostatic shocks \cite{Brainerd}. 
A Lorentz boost of the electron energy by a transformation from the jet 
frame into the Earth frame results in electron energies in the GeV-TeV 
range and could be further enhanced by the thermalization of the 
shock-reflected ion beam \cite{Wakefield,McClements2}. The shock-accelerated 
electrons undergo rapid velocity changes at the shock, and may radiate 
through synchrotron emission \cite{Synchrotron} and bremsstrahlung 
\cite{Schlickeiser}. The suppressed or delayed development of a reverse 
shock in our simulation may explain the absence of reverse shock signatures 
in GRB prompt emissions \cite{Lyutikov}.

{\bf Acknowledgments:} The DFG (grant SH-21/1-1) and the DIAS (Ireland) 
have funded this work. The Irish ICHEC and the Swedish HPC2N computer 
centres have provided computational facilities and support.

\end{document}